\documentclass[epj]{svjour}
\usepackage{graphicx}
\begin{document}
\newcommand{\ud}{\mathrm{d}}
\newcommand{\arcsec}{''}
\newcommand{\degree}{\mathrm{^o}}
\newcommand{\ion}[2]{#1$\,${\small #2}}
\title{Relativistic plasmas in AGN jets}
\subtitle{From synchrotron radiation to $\gamma$-ray emission}
\author{Giovanni La Mura\inst{1} \thanks{\emph{Present address:} giovanni.lamura@unipd.it} \and Giovanni Busetto\inst{1} \and Stefano Ciroi\inst{1} \and Piero Rafanelli\inst{1} \and Marco Berton\inst{1} \and Enrico Congiu\inst{1} \and Valentina Cracco\inst{1} \and Michele Frezzato\inst{1}
}                     
\institute{$^1$Dipartimento di Fisica e Astronomia - Universit\`a di Padova, Via F. Marzolo 8, 35131 Padova (Italy)}
\date{Received: date / Revised version: date}
%
\abstract{
Relativistic jets of plasma are a key ingredient of many types of Active Galactic Nuclei (AGN). Today we know that AGNs are powered by the accretion of inter stellar material into the gravitational field of a Super Massive Black Hole and that this process can release as much power as a whole galaxy, like the Milky Way, from a region that is comparable to the Solar System in size. Depending on the properties of the central energy source, a large fraction of this power can be involved in the acceleration of magnetized plasmas at relativistic speeds, to form large scale jets. The presence of jets affects the spectrum of AGNs through the emission of synchrotron radiation and Inverse Compton scattering of low energy photons, thus leading to a prominent non-thermal spectrum, some times extending from radio frequencies all the way up to $\gamma$-ray energies. Here we review some characteristic processes of radiation emission in AGN jets, which lead to the emission of photons in the radio, optical, X-ray and $\gamma$-ray bands, and we present the results of a spectroscopic campaign of optical counterparts. We discuss our observations and their connection with $\gamma$-ray properties in a scenario that traces the role of relativistic jets in different classes of AGNs, detected both in the local as well as in the remote Universe.
\PACS{
      {52.25.Os}{Emission, absorption and scattering of electromagnetic radiation}   \and
      {52.27.Ny}{Relativistic plasmas} \and
      {98.54.Cm}{Active and peculiar galaxies and related systems (including BL Lacertae objects, blazars, Seyfert galaxies, Markarian galaxies, and active galactic nuclei)} \and
      {98.70.Rz}{$\gamma$-ray sources; $\gamma$-ray bursts}
     } 
} 

\maketitle
\section{Introduction}
\label{secIntro}
After it was established that Active Galactic Nuclei (AGN) were exceptionally bright sources, with radiating powers ranging from $10^{42}\, {\rm erg\, s}^{-1}$ up to $10^{48}\, {\rm erg\, s}^{-1}$, often outshining the luminosities of entire galaxies, the interest of astrophysicists toward their spectra quickly grew and covered more and more extended frequencies of the electro-magnetic radiation. While the first extensive surveys of the sky in the radio domain discovered that some AGNs were characterized by very strong radio emission, pioneering observations carried out with high energy detectors outside the atmosphere identified many of them as X-ray sources \cite{Rothschild83,Mushotzky84,Turner89}. With the development of more sensitive and precise instruments, it became clear that AGNs are commonly characterized by radio and high energy properties that are unusual for the typical stellar processes of quiescent galaxies.

Although there are remarkable differences between the properties of various types of AGNs, several attempts to reconstruct their intrinsic spectra over the whole range of accessible radiation frequencies were made \cite{Elvis94,Ho99,Richards06,Shang11,Krawczyk13}. Today we know that AGNs can radiate their power from radio frequencies all the way up to the high energy domain of TeV $\gamma$-rays. The idea that non-thermal processes, involving relativistic plasmas in intense magnetic fields, could explain the emission of radiation over several orders of magnitude in frequency became soon very popular \cite{Lovelace76,Blandford76}. Adopting the AGN unified model, where we assume that the energy source of the central engine is accretion of matter into the gravitational field of a Super Massive Black Hole (SMBH, $10^6\, {\rm M}_\odot \leq M_{BH} < 10^{10}\, {\rm M}_\odot$), we naturally expect the presence of a relativistic accretion flow, which must consist of hot plasma, threaded by strong magnetic fields. In these conditions, a combination of intense, mostly anisotropic, thermal and non-thermal radiation components must be produced \cite{Blandford77,Blandford82} and it is eventually observed with different characteristics, depending on the orientation of the source, the amount of obscuration and the rate of the accretion process itself \cite{Antonucci93,Heckman14}.

While this overall picture can provide a fairly satisfactory explanation of many AGN observations, the theoretical details involve understanding of advanced physical concepts. A particularly interesting case arises when the accretion flow couples with the magnetic field in such a way that extremely well collimated jets of plasma are accelerated away from the nucleus at relativistic speeds. Many AGNs, when observed at high angular resolutions, show evidence of such relativistic jets \cite{Oshlack01,Jorstad05}. The jets carry a significant fraction of the AGN power away from the nucleus and they are observed to extend from the sub-parsec scale, out to several kilo-parsecs, or even to the mega-parsec scale. Their main radiation mechanism involves synchrotron emission, coupled, in some sources, with inverse Compton scattering of photons onto highly relativistic charged particles. The jets can be either modestly developed or they can represent a dominant feature of the AGN. Thanks to the unprecedented sensitivity and resolution achieved by the {\it Large Area Telescope} on the {\it Fermi} satellite (Fermi/LAT, \cite{Atwood09}) at $\gamma$-ray energies between $100\,$MeV and $300\,$GeV, it turned out that most of the extra-galactic radiation sources in this spectral window are actually AGNs with powerful jets, populating both the local and the remote Universe \cite{Ackermann15}. In spite of the advances in our understanding of AGNs, however, many fundamental questions, concerning the jet composition, their acceleration and collimation processes, as well as the origin of their radiation itself, are still open.

In this work, we present a review of some fundamental aspects of jet radiation theory and we discuss it in view of the most recent achievements of AGN jet investigations with high energy detectors and multiple frequency counterpart analysis. We use the latest available compilation of $\gamma$-ray sources detected by Fermi/LAT to search for AGNs with strong jet components and we seek coincident X-ray and radio counterparts to constrain the likely position of the high energy sources. We finally describe the results of our ongoing campaign to obtain optical spectra of the counterparts, discussing the role of jet activity in different AGN families and the implied constraints on the physics of jets.

\section{Theory of AGN jet radiation}
\label{secTheory}
The detection of jets from galactic nuclei traces back to the earliest epochs of morphological studies \cite{Curtis18}, but evidence for relativistic motions primarily came from high resolution interferometric radio observations of radio loud QSOs. Relativistic plasmas moving in strong magnetic fields, through a comparatively static medium, must produce electromagnetic radiation, shock fronts and be subject to several types of instabilities. Some of these instabilities are suppressed by large bulk Lorentz factors or by assuming multi-component jet kinematic structures. Both features are apparently well supported by observations. An exhaustive review of the theory of AGN jets and the models that are supposed to stabilize the jet structure and to provide large scale coherence can be found in \cite{RelAGNJets}. For what concerns our interests, limited to the treatment of the jet radiation mechanisms and to the investigation of jet distribution in different types of AGNs by means of multiple wavelength analysis and optical observations, we will recall only the most relevant fundamentals.

If we consider a single electrically charged particle, moving at relativistic speed $\beta = v / c$ in a magnetic field of energy density $u_B$, we expect a synchrotron energy loss rate that we can express in terms of the particle Lorentz factor $\Gamma_p$:
$$\left( \frac{\ud \Gamma_p}{\ud t} \right)_{sy} = - \frac{4}{3} c \sigma_{T} \frac{u_B}{m_e c^2} Z^4 \left( \frac{m_e}{m} \right)^3 \beta^2 \Gamma_p^2, \eqno(1)$$
where $\sigma_{T}$ is the Thomson cross-section, $Z$ is the particle charge in units of the electron charge $e$, $m$ its rest frame mass, while $m_e$ is the electron mass. If the moving particle is an electron, most of the power is radiated close to a characteristic frequency:
$$\nu_c = 3 \Gamma_p^2 \frac{e B \sin \psi}{4 \pi m_e c}, \eqno(2)$$
where $B$ is the magnetic field intensity, $e$ is the fundamental electron charge, while $\psi$ represents the particle motion's pitch angle, with respect to the local magnetic field direction. An estimate of the total power radiated by a population of electrons, with an initial random velocity distribution, can be derived from Eq.~(1) by averaging among all possible directions and using the so called $\delta$-approximation, where we assume that all the power is radiated at $\nu_c$:
$$P_\nu^{sy,\delta}(\Gamma_p) = \frac{32 \pi}{9} \left( \frac{e^2}{m_e c^2} \right)^2 u_B \beta^2 \Gamma_p^2 \delta(\nu - \nu_c), \eqno(3)$$
with $\delta(\nu - \nu_c)$ representing a Dirac's $\delta$-function centered at $\nu_c$. The synchrotron emission coefficient is obtained by integrating the particle energy distribution with the energy loss rate of Eq.~(3) over all the possible energies:
$$j_\nu^{sy} = \frac{1}{4 \pi} \int_1^\infty \ud \Gamma_p n(\Gamma_p) P_\nu^{sy,\delta}(\Gamma_p), \eqno(4)$$
where we introduced $n(\Gamma_p)$, the particle number density distribution as a function of $\Gamma_p$. Adopting a truncated power-law distribution of charged particles:
\begin{center}
  $n(\Gamma_p) = k \Gamma_p^{-p}\;$ for $\; \Gamma_{p, min} \leq \Gamma_p \leq \Gamma_{p, max}$
\end{center}
Eq.~(4), combined with Eq.~(2), predicts a power-law emission coefficient $j_\nu \propto \nu^{(p - 1) / 2}$ \cite{Rybicki91}. In optically thin plasmas, the derived expression of $j_\nu$ results in the emission of a power-law spectrum, having slope controlled by the energy distribution of the radiating particles. Optically thick material, where synchrotron self absorption assumes a relevant role, leads to a spectrum of theoretical intensity $I_\nu \propto \nu^{5 / 2}$ below a turn-off frequency, while, above it, the optically thin regime is restored and $I_\nu \propto j_\nu$. The combined emission of a homogeneous population of radiating particles is characterized by a high degree of linear polarization, with the direction of polarization controlled by the dominant magnetic field component.

In addition to the synchrotron emission, the mixture of highly relativistic particles with a radiation field, where the average photon energies lie below the particle total energies, leads to the rise of significant Inverse Compton (IC) scattering processes. The general theory of such events can be well understood, again starting from the single particle case, by evaluating the process in the particle's rest frame, where it appears as a regular Compton scattering. The full treatment of a radiation field scattered by a population of relativistic charged particles involves integration of Klein-Nishina cross-sections and non trivial relativistic corrections. It is possible, however, to derive approximate expressions of the radiation emission coefficient, in analogy to the previous $\delta$-approximation, assuming that all the scatterings lead to emission of radiation, whose frequency only depends on the original photon frequency and on the scattering particle's energy:
$$j_\nu^{IC,\delta} = \frac{h c \sigma_{T} \epsilon_S^2}{8 \pi} \int_{\epsilon_S}^\infty \ud \Gamma_p \frac{n_e(\Gamma_p)}{\Gamma_p^4} \int_{\epsilon_S / (2 \Gamma_p^2)}^\infty \ud \epsilon \frac{n_{ph}(\epsilon)}{\epsilon^2}, \eqno(5)$$
where $\epsilon$ and $\epsilon_S$ are, respectively, the energy of the photon before and after scattering, normalized to the rest mass energy of the electron:
$$\epsilon_{(S)} = \frac{h \nu_{(S)}}{m_e c^2}. \eqno(6)$$

Provided a high enough particle energy, it turns out that the scattering processes effectively transfers energy to the photons, which emerge with an average frequency increase that is roughly proportional to $\Gamma^2$ ($\Gamma$ being the bulk Lorentz factor of the plasma). This increase, combined with relativistic aberration and boosting effects, which lead to observed flux densities related to the intrinsic ones by:
$$F_{\nu^{obs}}^{obs} = d^3 F_{\nu^{em}}^{em}, \eqno(7)$$
where
$$d = [ \Gamma ( 1 - \beta_{\Gamma} \mu^{obs})]^{-1} \eqno(8)$$
is the Doppler factor, $F_{\nu^{obs}}^{obs}$ is the observed flux in the observer's rest frame measured frequency, $F_{\nu^{em}}^{em}$ is the emitted flux in the emission frame frequency and $\mu^{obs}$ is the cosine of the angle between the line of sight and the plasma moving direction, results in the production of extremely energetic photons, propagating in close alignment with the plasma direction of motion.

Another source of radiation emission, which accounts for the production of high energy photons in a broader beam, is the interaction of different kinematic layers in a complex jet structure. In the so called spine-sheath scenario, where a ultra-relativistic jet spine is coated by a mildly relativistic sheath that supports large scale stability, the shear between the different components leads to the development of shock fronts and violent acceleration sites. The energy exchanged in such regions is sufficiently high to activate hadronic cascades, with a substantial production of unstable particle species that later decay into high energy photons.

\begin{figure}[t]
  \begin{center}
    \includegraphics[width=0.48\textwidth]{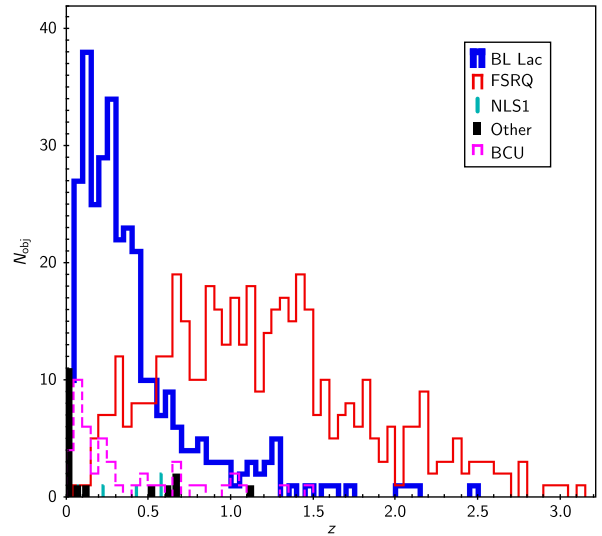}
  \end{center}
  \caption{The redshift distribution of the Fermi/LAT detected AGNs with known redshift, according to the 3$^{rd}$ LAT AGN Catalogue (3LAC \cite{Ackermann15}). Bar styles are used to outline the distribution of different AGN types: thick continuous line for BL Lacs; thin continuous line for FSRQs; thin spikes for NLS1s; thin dashed line for BCUs and thick spikes for other types of AGNs. \label{3LAC_Histo}}
\end{figure}
\section{AGN jet demographics}
\label{secData}
Based on the results summarized in Eqs.~(7-8), we expect that the jet radiation is most prominent in the case of highly relativistic motion ($\Gamma >> 1$, $\beta_\Gamma \approx 1$) occurring close to the line of sight ($\mu^{obs} \approx 1$). Both conditions apply to the family of AGNs named {\it blazars}. These objects comprise a class of radio loud AGNs, characterized by flat radio spectra, high brightness temperatures ($T_b \geq 10^9\,$K), remarkable variability (below the hour time scale for the highest energy emitting sources) and high degree of polarization ($\sim 3\%$). From an optical point of view, blazar spectra can be further divided in two categories: Flat Spectrum Radio Quasars (FSRQ), with prominent broad emission lines, and BL Lac type objects (BLL), with only a power-law continuum, possibly weak emission lines (having equivalent widths $EW \leq 5\,$\AA) and, sometimes, a host galaxy contribution with absorption lines. Due to the relativistic boosting of their jet radiation, which enhances their visibility at large distances, these two families of blazars, together with a conspicuous fraction of {\it blazar candidates of uncertain type} (BCU), represent the dominant sources of extra-galactic high energy radiation. This situation is fairly well illustrated in Fig.~\ref{3LAC_Histo}, where we plot the distribution of Fermi/LAT detected AGNs in the energy range $100\, \mathrm{MeV} \leq E_\gamma \leq 300\, \mathrm{GeV}$, according to their classification and their redshift, when available. Even if the redshift distributions of the BLL and BCU classes are severely incomplete, due to the lack of firm identification of spectral features for many BLLs and of published high quality spectra for most BCUs, the dominance of blazars over lower power or possibly misaligned jet sources is obvious. Although non-blazar AGN classes, such as radio-galaxies (RDG), Soft Spectrum Radio Quasars (SSRQ) and other types of AGNs listed in the 3LAC, are known to host jets, they just account for a minor fraction of the $\gamma$-ray sources, and, but for TXS~0348+013 (SSRQ at $z = 1.12$), their $\gamma$-ray detection has only been possible up to $z < 0.7$.

\begin{figure*}[t]
  \begin{center}
    \includegraphics[width=\textwidth]{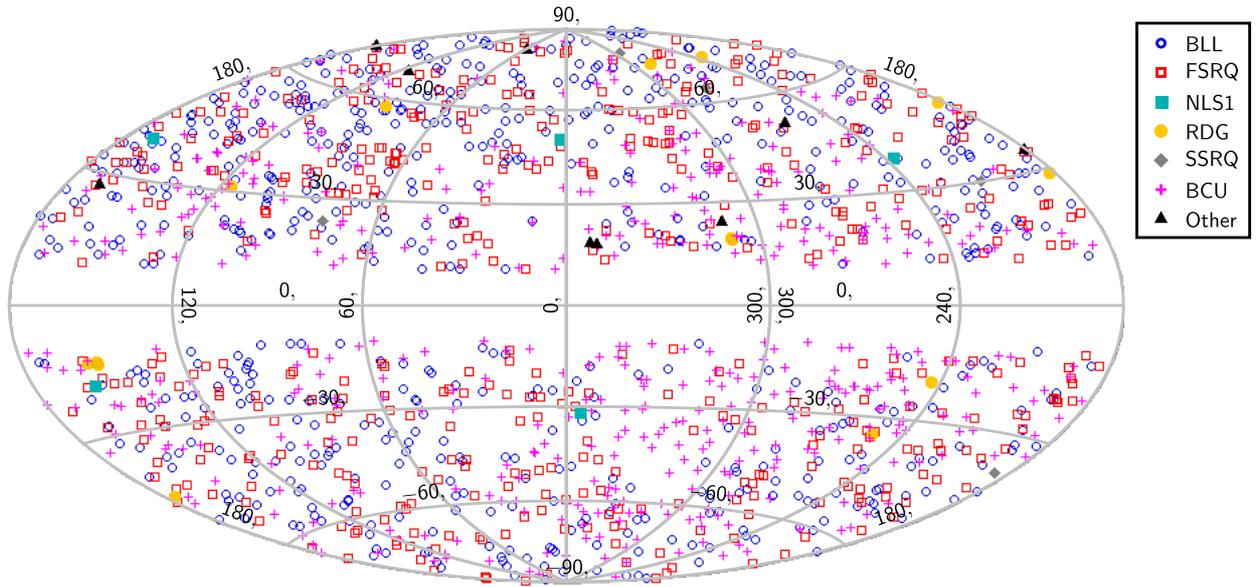}
  \end{center}
  \vspace{-1cm}
  \caption{Map of the 3LAC detected AGNs in galactic coordinates. The symbols mark different types of objects: empty circles for BL Lacs; empty squares for FSRQs, filled squares for NLS1s; filled circles for RDGs; filled diamonds for SSRQs; crosses for BCUs and filled triangles for other types of AGNs. The rarest $\gamma$-ray emitting AGN classes have been indicated with larger markers. The high background emission and the severe extinction of lower energy counterparts close to the Galactic plane forced to ignore objects with galactic latitude $|b| \leq 10\degree$. \label{3LAC_Sources}}
\end{figure*}
A special case is represented by the class of $\gamma$-ray emitting Narrow Line Seyfert 1s (NLS1). These are AGN characterized by the spectrum of Seyfert 1 galaxies, but with a broad line Doppler width corresponding to a velocity field smaller than $2000\, \mathrm{km\, s}^{-1}$ in the H$_\beta$ emission line at $\lambda_0 = 4861\,$\AA. They are commonly interpreted as young AGNs, with a relatively small mass black hole ($M_{BH} \leq 10^8\, \mathrm{M}_\odot$) that is actively growing \cite{Foschini15,Berton16}. They are, therefore, analogous to FSRQs, although generally less luminous, but they can still be detected, like for example in the case of 3FGL~J1222.4+0414, first classified as FSRQ at $z = 0.966$, due to its prominent \ion{Mg}{II} $\lambda2798$ line, but then identified as NLS1, thanks to the Sloan Digital Sky Survey Baryon Oscillation Spectroscopy Survey (SDSS-BOSS \cite{Smee13,Yao15}) spectrum extending far enough into the IR to observe the profile of its H$_\beta$.

The illustrated picture is generally well understood in terms of the highly anisotropic emission of radiation from AGN jets. If most of the radiation propagates in a narrow beam that is closely aligned with the source direction of motion, only the objects with jets aligned with the line of sight will be bright enough to be observed at large distances. As soon as the alignment is lost, the observed jet radiation becomes so weak that it can only be detected in nearby objects. A full understanding of jet physics, however, can only be drawn from a comprehensive analysis of the intrinsic jet power, since we can expect that a large number of sources hosting weaker jets than the classical blazars can exist and that they can play a relevant role in the explanation of the AGN contribution to the high energy background radiation field.

\section{Multiple-wavelength investigations and optical follow-ups}
\label{secObs}
Thanks to the execution of large survey programs, which in the last decades have been regularly monitoring the sky across the whole range of the electro-magnetic spectrum, we are currently able to apply investigation techniques that identify AGN activity with unprecedented efficiency. Based on the strong non-thermal spectra, produced by sources with powerful relativistic jets, \cite{Massaro11,DAbrusco12,Massaro12} identified a sequence of infra-red colors from the Wide-field Infrared Survey Explorer (WISE) that specifically characterizes blazars, especially in the case of strong $\gamma$-ray emission. The execution of spectroscopic campaigns, such as the SDSS \cite{York00,Albareti16}, or the targeted optical spectroscopic investigations led by \cite{Alvarez16a,Alvarez16b,Alvarez16c,Massaro16} on the basis of the proposed counterparts to high energy sources, mostly confirmed the blazar nature of several candidates, providing, in some cases, additional information of the source redshifts and intrinsic powers. Combining the first 4 years of monitoring with the Fermi/LAT telescope with spectroscopic follow-ups, the resulting distribution of known AGNs, emitting $\gamma$-ray radiation, turned out to appear as it is illustrated in Fig.~\ref{3LAC_Sources}.

As it can be appreciated in Fig.~\ref{3LAC_Sources}, in spite of the extensive efforts to characterize the multiple-frequency properties of the counterparts to $\gamma$-ray AGNs, many objects still lack a clear classification. Following a different approach, based on the analysis of the $\gamma$-ray variability properties, \cite{Chiaro16} proposed a machine-learning procedure that is able to predict the classification likelihood of a large fraction of the still uncertain sources. The method adopted by \cite{Chiaro16} has been paired with an optical spectroscopic campaign, carried out at the Asiago Astrophysical Observatory, supported by a selection of potential counterparts to the $\gamma$-ray sources that is based on the combination of radio and X-ray observations. Since the high energy $\gamma$-rays cannot be focused by optical systems, indeed, the reconstruction of their incidence angle and, therefore, the correct location of the source on the sky, can only be modeled with a precision that, in optimal conditions, only reduces to a few arc minutes. Assuming that the $\gamma$-ray emission of the detected objects arises from relativistic jets, however, we expect a significant flux to be produced at X-ray and radio wavelengths, due to the synchrotron energy losses expected for a population of relativistic particles that scatter photons at $\gamma$-ray energies. The angular resolutions of detectors working at these frequencies provide a much better localization of the source.

\begin{table}
  \caption{Instrumental characteristics of the Asiago Astrophysical Observatory Telescopes. \label{tabAsiago}}
  \begin{center}
    \begin{tabular}{lc}
      \hline
      \hline
      {\bf Copernico Telescope}$^{\rm a}$ & \\
      \hline
      Main mirror diameter & $1.82\,$m \\
      Focal length & $5.39\,$m \\
      Spectrograph & AFOSC$^{\rm b}$ \\
      Entrance slit width & $1.69\arcsec$ \\
      Grating & $300\,$gr mm$^{-1}$ \\
      Wavelength range & $3700$--$8000\,$\AA \\
      Spectral resolution & $600$ \\
      \hline
      {\bf Galileo Telescope}$^{\rm c}$ & \\
      \hline
      Main mirror diameter & $1.22\,$m \\
      Focal length & $6.00\,$m \\
      Spectrograph & Boller \& Chivens \\
      Entrance slit width & $3.5 - 5.0\arcsec$ \\
      Grating & $300\,$gr mm$^{-1}$ \\
      Wavelength range & $3500$--$7500\,$\AA \\
      Spectral resolution & $600$ \\
      \hline
    \end{tabular}
  \end{center}
      {\footnotesize
        $^{\rm a}$Website: \texttt{http://www.oapd.inaf.it/index.php/en} \\
        $^{\rm b}$Asiago Faint Object Spectrograph and Camera \\
        $^{\rm c}$Website: \texttt{http://www.dfa.unipd.it/index.php?id=300}
        }
\end{table}
By matching X-ray and radio sources with a positional uncertainty tolerance of 10\arcsec, in order to account for the lower accuracy of extensive X-ray surveys and off-axis detections of targeted observations, we are generally able to narrow down the number of potential counterparts to the $\gamma$-ray sources, which, if associated with optical sources having visual magnitude $V \leq 18$, are subsequently observed in long slit spectroscopy. By combining observations of the selected targets with regular exposures of spectro-photometric standards and calibration lamps, we are able to extract optical spectra with resolution $R \sim 600$ on a wavelength range running approximately from 3700\AA\ up to 8000\AA\ (the exact limits depending on the telescopes and instruments, whose main characteristics are listed in Table~\ref{tabAsiago}). An average exposure time of 2 hours is generally required to collect spectra with SNR$\, \sim 20$ in the continuum of sources with $V = 17$. Using standard IRAF\footnote{IRAF is the {\it Image Reduction and Analysis Facility} distributed at: \texttt{http://iraf.noao.edu}} tasks to extract mono-dimensional spectra, the data are subsequently analyzed, in order to characterize the shape of the continuum radiation and to identify possible emission and absorption lines, which place constraints on the classification of the source and may lead to an estimate of its redshift.

The observational campaign of unclassified sources is further paired with a monitoring program of bright AGNs hosting powerful jets, which is specifically triggered in the case of significant outbursts of the sources. The aim of such additional observations is to improve the statistics of our current knowledge of the multiple-wavelength variability properties in well studied AGN jets, both to better constrain the source of radiation outbursts, as well as to identify similar footprints in the light curves and spectral properties of unclassified objects.

\section{Optical spectra from AGNs with jets}
\label{secResults}
Systematic differences in the optical spectra of radio loud sources, likely hosting a relativistic jet, with respect to radio quiet objects have long been known \cite{Osterbrock06}. These suggest that a dynamic interplay between the jet plasma and the surrounding environment takes place at the smallest sub-parsec scales. Moreover, the relativistic jets produced by AGNs contribute to the radiated spectrum with a non-thermal component that becomes dominant in the case of blazars. The low energy part of the spectrum, from radio wavelengths up to X-ray frequencies, is well explained in terms of synchrotron emission from relativistic charged particles. The high energy part, which often appears as a second peak in the $\gamma$-ray domain of the Spectral Energy Distribution (SED), instead, is interpreted as the result of low energy photons scattered off the ultra-relativistic particles of the jet and, therefore, propagating mostly in the direction of the jet itself. This scenario accounts very well for the observational evidence that blazars are the most frequent class of AGNs detected in high energy at large redshift. It also predicts that the optical spectrum of $\gamma$-ray detected AGNs should be dominated by the jet contribution in nearly all, but the closest objects. The origin of the scattered photons is not completely clear, since they can be synchrotron photons of the jet itself, as well as low energy photons produced in other sites of the AGN, such as its accretion disk, its Broad Line Region (BLR), the distribution of dusty molecular gas that surrounds the central source and is heated by its intense radiation field, or even Cosmic Microwave Background (CMB) photons. The location of the $\gamma$-ray production site itself is not completely understood, with some studies suggesting an effective production close to the SMBH, well within the BLR \cite{Tavecchio10,Poutanen10}, and others pointing to a much farther site, up to several parsecs away \cite{Sikora08,Jorstad10}. In the case of 3C~454.3, a powerful $\gamma$-ray FSRQ, a significant increase of the broad \ion{Mg}{II} $\lambda2798$ has been reported in response to $\gamma$-ray flaring activity, suggesting a close spatial connection \cite{LeonTavares13}. Some rest frame optical lines like H$\beta$ and H$\gamma$, though variable, are consistent with longer time scales with respect to the characteristic flaring time \cite{Isler13}. Recently, the conditions under which we can expect significant optical depth for photon interactions that might lead to substantial absorptions of $\gamma$-rays, if the source lies within the BLR, have been modeled by means of detailed photoionization calculations, suggesting that the breaks observed in the $\gamma$-ray spectra of some FSRQs and Low Synchrotron Peak (LSP) BL Lacs could arise as the result of $\gamma$-ray emission from within the BLR \cite{Abolmasov17}.

Optical investigations carried out by means of public spectroscopic surveys, such as the SDSS, or through specific observations at the telescope, tend to agree with the prediction that $\gamma$-ray emission is generally associated with dominant non-thermal spectra at every wavelength. A comprehensive investigation on this matter is subject to the difficulties connected with the associations of high energy sources with their lower frequency counterparts. Although in many cases we can relay on the idea that a jet powered $\gamma$-ray source will be associated with strong enough radio and X-ray counterparts, which narrow down the number of potential optical associations in a typical $\gamma$-ray signal confinement area, the technique does not apply everywhere with the same efficiency. In addition to the effects of extinction from the interstellar medium (ISM) close to the plane of the Milky Way, which may hide the source from optical observations, large column densities of neutral gas are very likely to absorb X-ray photons, making the counterpart identification more difficult.

\begin{figure}[t]
  \begin{center}
    \includegraphics[width=0.48\textwidth]{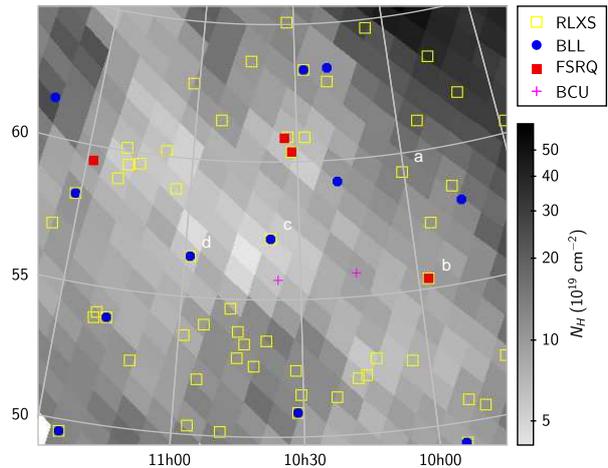}
  \end{center}
  \caption{The location of coincident X-ray/radio sources extracted from the ROSAT and the NVSS survey catalogs (large empty boxes) in the area around the Lockman Hole, overlaid to the Galactic neutral Hydrogen column density map (\cite{Kalberla05}, grey scale), plotted in equatorial coordinates (J2000). 3LAC $\gamma$-ray sources are denoted as filled circles for BL Lacs, filled squares for FSRQs and crosses for BCUs. Labels correspond to the optical spectra shown in the panels of Fig.~\ref{Spec_SDSS}. \label{Lockman_Map}}
\end{figure}
In principle, the best way to investigate the intrinsic properties of sources outside the Galaxy is to analyze objects located in regions of the sky with minimal effects of radiation absorption. An example of such a region is illustrated in Fig.~\ref{Lockman_Map}, where the neutral Hydrogen column density map, derived from the Leiden-Argentine-Bonn radio observations (LAB \cite{Kalberla05}), is overlaid with the distribution of coincident radio and X-ray sources extracted from the {\it National radio astronomy observatory Very large array Sky Survey} (NVSS, \cite{Condon98}) and the {\it R\"ontgen Satellite} survey (ROSAT, \cite{Boller16}) in the area surrounding the {\it Lockman Hole}. This is a specific region of the sky, where we detect the minimum amount of neutral gas column density from the Galaxy ($N_H \approx 5 \cdot 10^{19}\, \mathrm{cm}^{-2}$, \cite{Lockman86,Dickey90}). Due to their flux limited sensitivity, these surveys well represent the effects of absorptions on the source detection distribution and, as Fig.~\ref{Lockman_Map} shows, there is a strong trend for classified $\gamma$-ray sources to be coincident or to lie very close to corresponding radio-loud X-ray sources (RLXS). Sources without a firm classification, on the other hand, lack a confident association with the lower energy counterparts that should be observed to extract the spectrum.

\begin{figure*}[t]
  \begin{center}
    \includegraphics[width=0.48\textwidth]{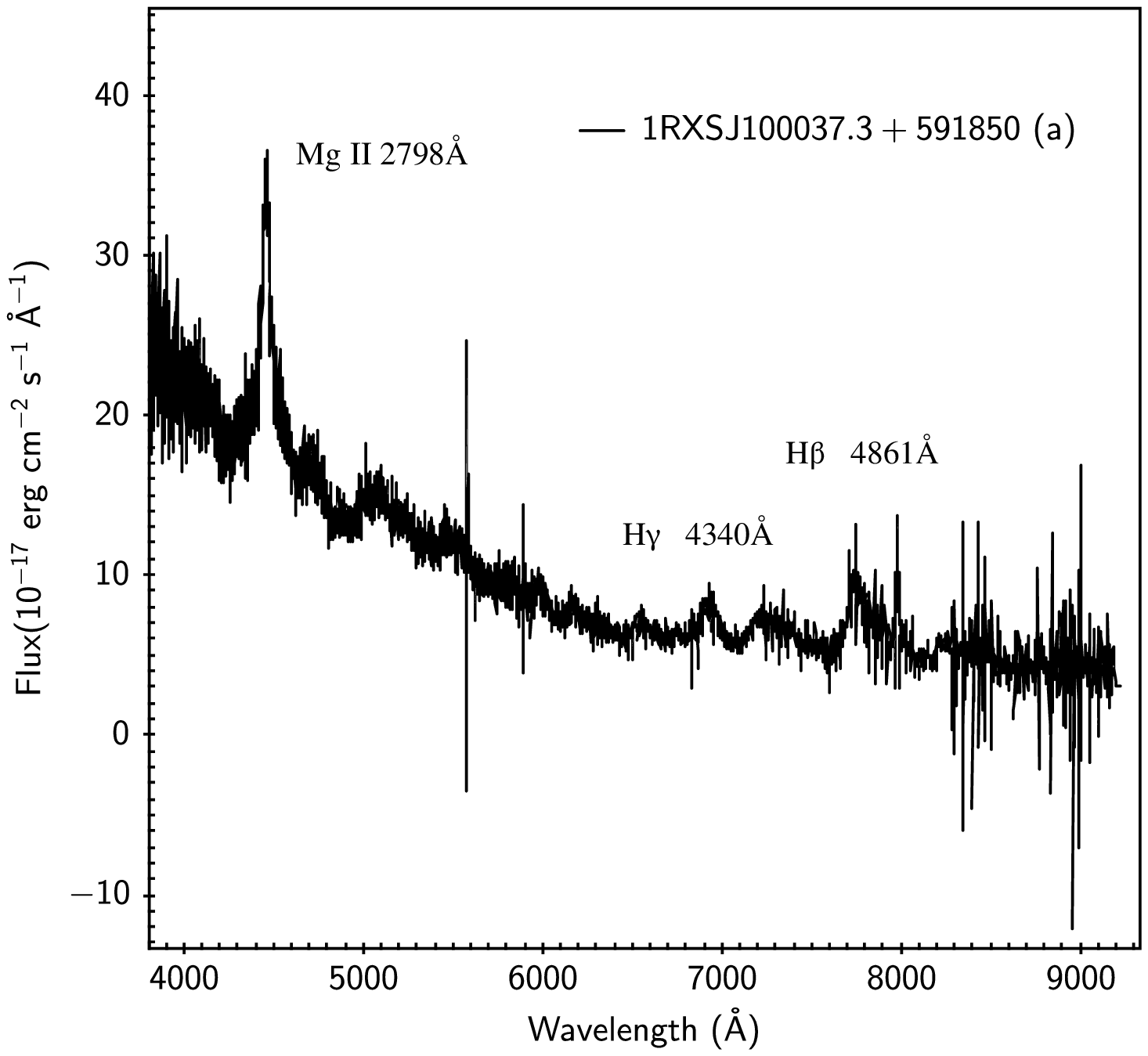}
    \includegraphics[width=0.48\textwidth]{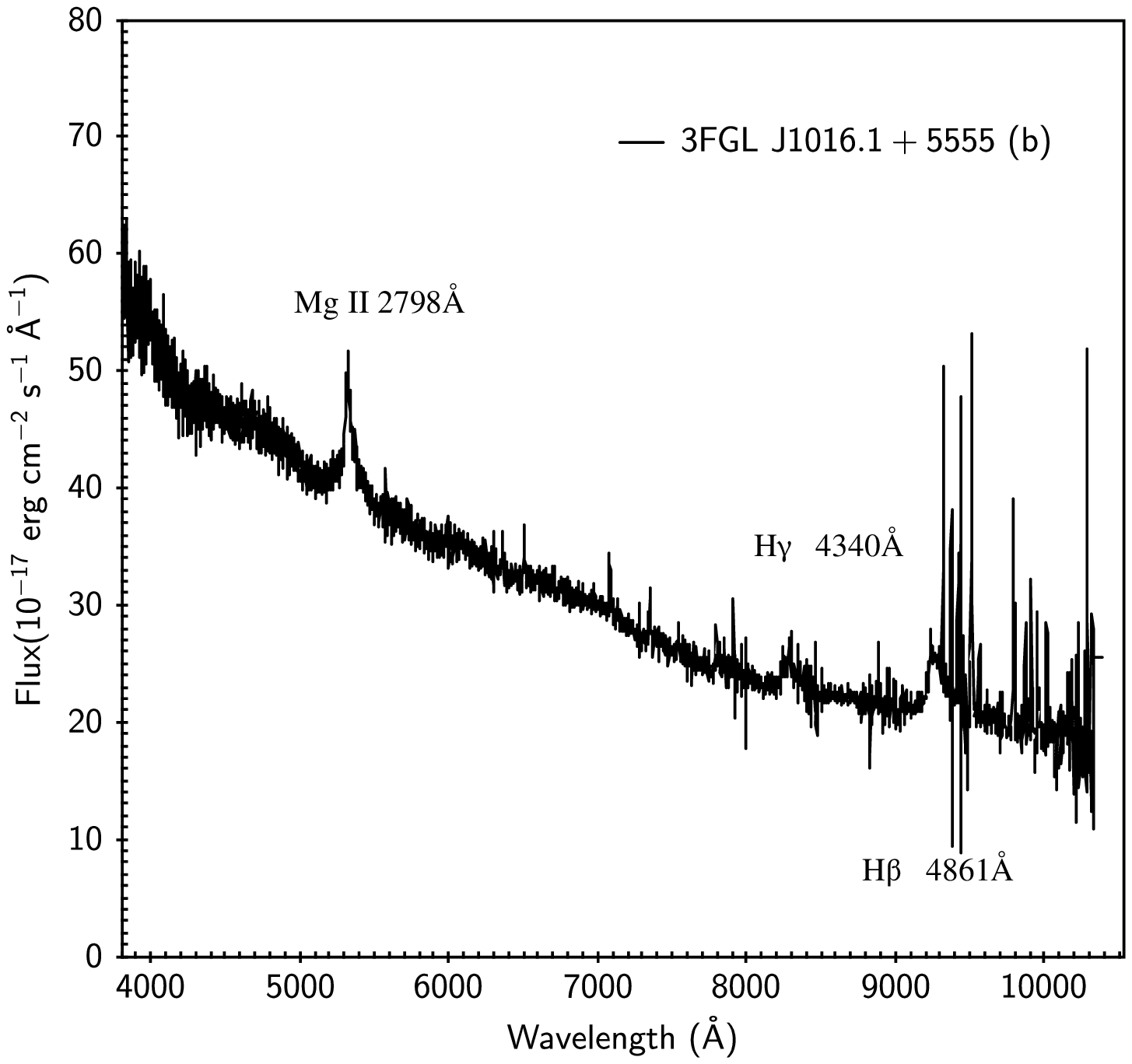}
    \includegraphics[width=0.48\textwidth]{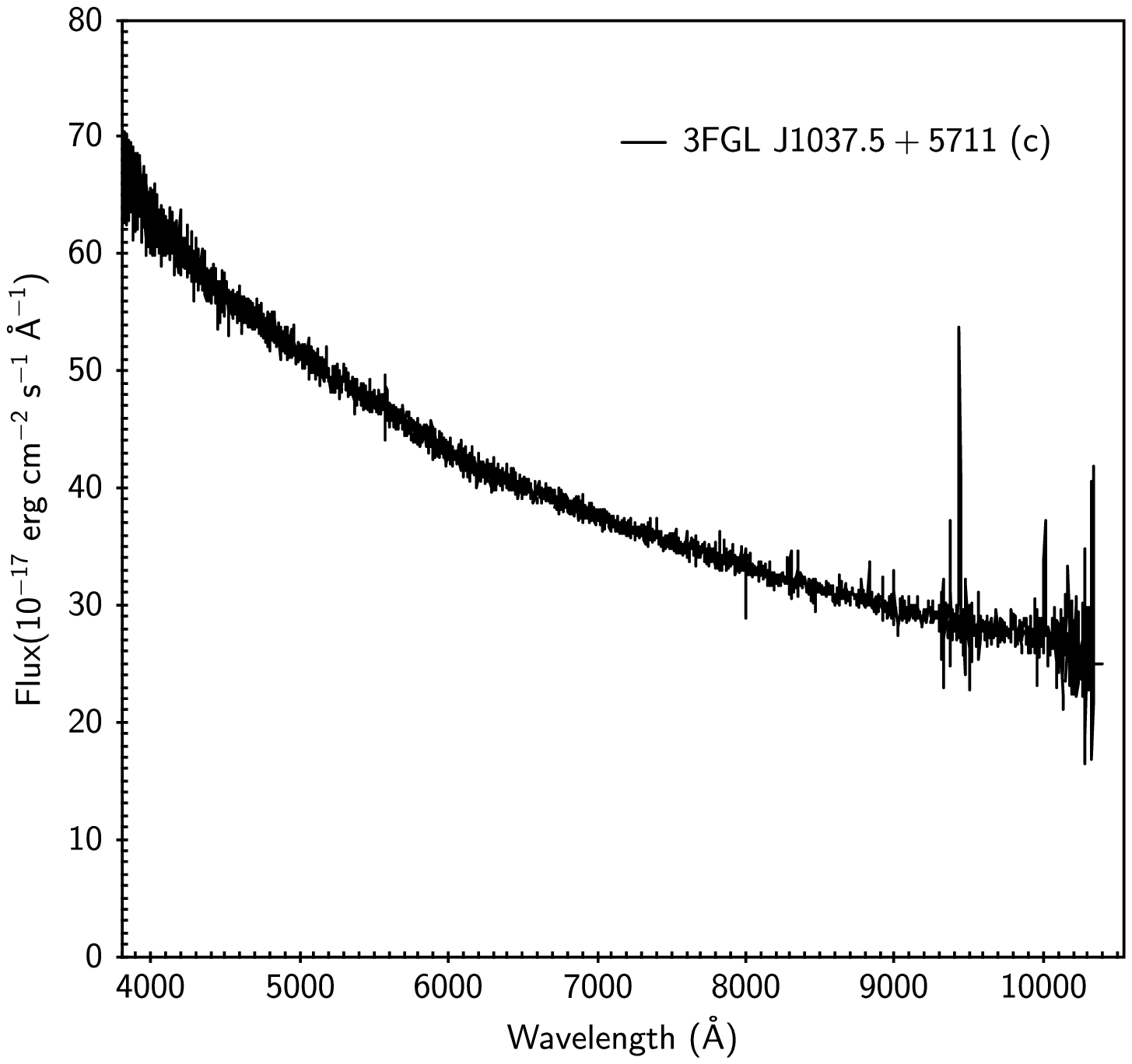}
    \includegraphics[width=0.48\textwidth]{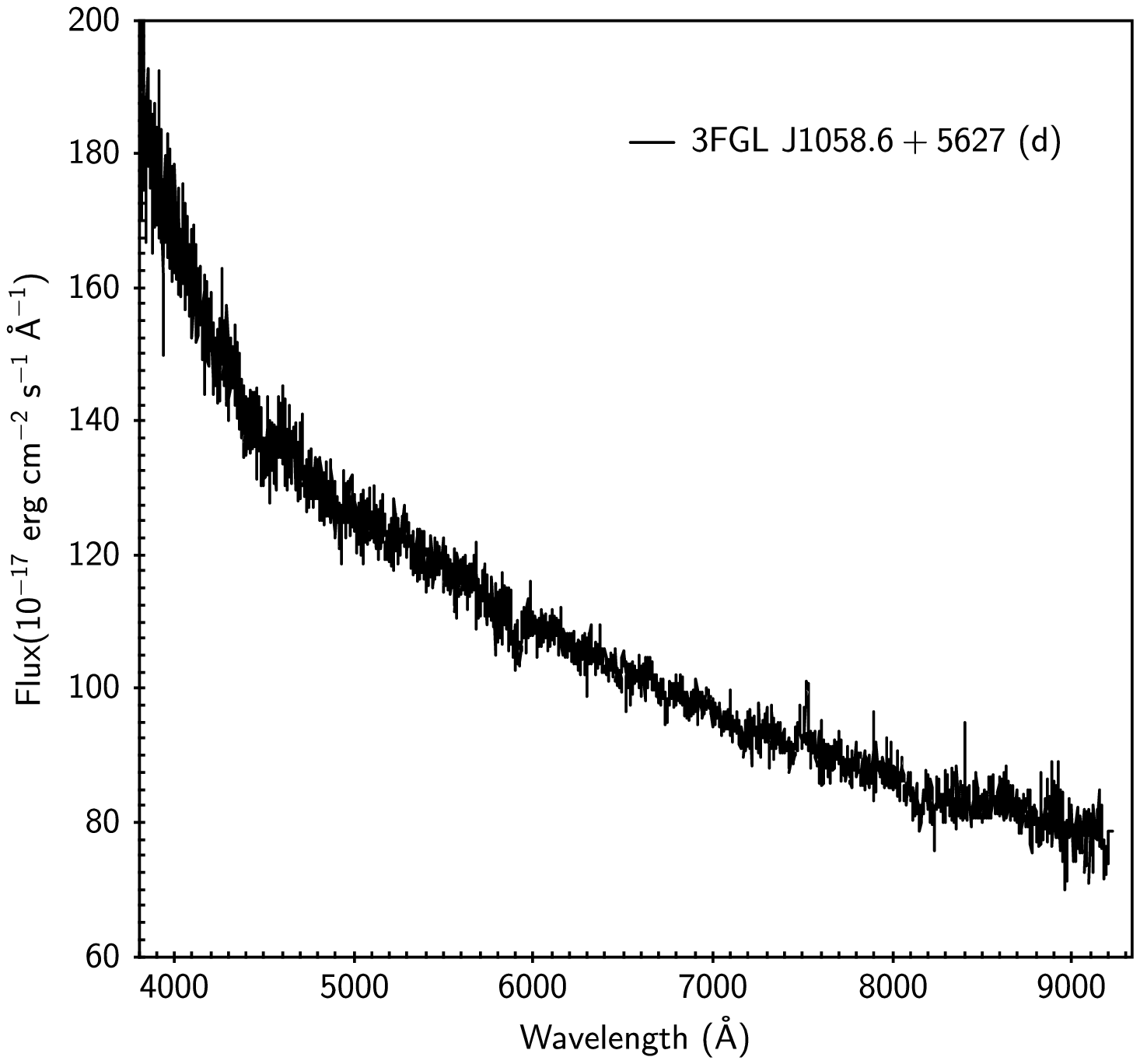}
  \end{center}
  \caption{Example SDSS spectra from the labeled sources in the Lockman Hole region. (a) 1RXS~J100037.3+591850 is a $\gamma$-ray undetected QSO, located at $z = 0.594$, with prominent broad emission lines and flat radio spectrum. (b) 3FGL~J1016.1+5555 is a FSRQ located at $z = 0.899$ and detected as $\gamma$-ray source. (c, d) 3FGL~J1016.1+5555 and 3FGL~J1016.1+5555 are BL Lac type objects detected in the $\gamma$-rays, but without a firm redshift measurement. The $\gamma$-ray detected sources share a strong power law continuum, suggestive of dominant jet contribution to the optical spectra. The source in panel (a), on the contrary, has a clearer thermal upturn below $\lambda = 6300\,$\AA\ (rest frame $\lambda_0 = 3952\,$\AA), consistent with a comparatively weaker jet power and with its lack of high energy detection. \label{Spec_SDSS}}
\end{figure*}

Fig.~\ref{Spec_SDSS} illustrates a few example spectra of some RLXSs located in the Lockman Hole region, both for objects associated with $\gamma$-ray sources detected by the Fermi/LAT, as well as for objects without $\gamma$-ray detection (in the 4 year long Fermi monitoring campaign of 3LAC). Additional spectra, obtained in other sky areas during the blazar monitoring campaign, carried out at the Asiago Astrophysical Observatory, instead, are plotted in Fig.~\ref{Spec_Asiago}. As it is expected, in the vast majority of cases, the optical spectra show a clear non-thermal power-law continuum, though with a variety of different slopes, sometimes featureless and some other times accompanied by the prominent broad emission lines of AGNs. By inspecting the spectra of objects associated with $\gamma$-ray emission, we recognize the characteristic signature of a power law continuum in all sources located at $z \geq 0.1$ and, sometimes, even in closer objects. This non-thermal component gets mixed with other spectral contributions, like the thermal continuum in 1RXS~J100037.3 +591850, or the host galaxy integrated star light in some BL Lac type objects, such as the BCU 3FGL~J0602.2+5314, which are strongly suggestive of a comparatively weaker intrinsic jet activity.

\section{Result discussion and implications}
\label{secDiscussion}
Optical spectroscopic investigations of jet powered AGNs have a fundamental importance for high energy Astrophysics. From a cosmological point of view, blazars and other types of powerful AGNs probe the history of cosmological structure formation, they measure the opacity of the Universe for photons of various energies and probably they played a primary role in affecting the status of the inter galactic medium. Unfortunately, our ability to constrain fundamental cosmological parameters on the basis of AGN observations is limited by severe uncertainties on the intrinsic power of such objects. The main factors of indetermination that we have to deal with, when working on AGNs, descend from the still poorly constrained degree of anisotropy in the different radiation components used to estimate the luminosity, from the uncertainty in the power emitted at different frequencies and, in many cases, from the unknown distance.

Optical spectroscopy has always been considered a privileged investigation technique, to reduce the above mentioned uncertainty sources, because it provides a consistent classification scheme of the objects that we can use to infer relationships between different spectral windows. In addition, it provides the best estimates of distance, on a cosmological scale, through the measurement of redshifts. Both of these advantages, however, relay on the identification of well defined spectral features (mainly emission lines), which can be confidently related to the source. In the case of blazars, and in particular for BL Lac type objects and BCUs, these features are most frequently missing.

\begin{figure*}[t]
  \begin{center}
    \includegraphics[width=0.48\textwidth]{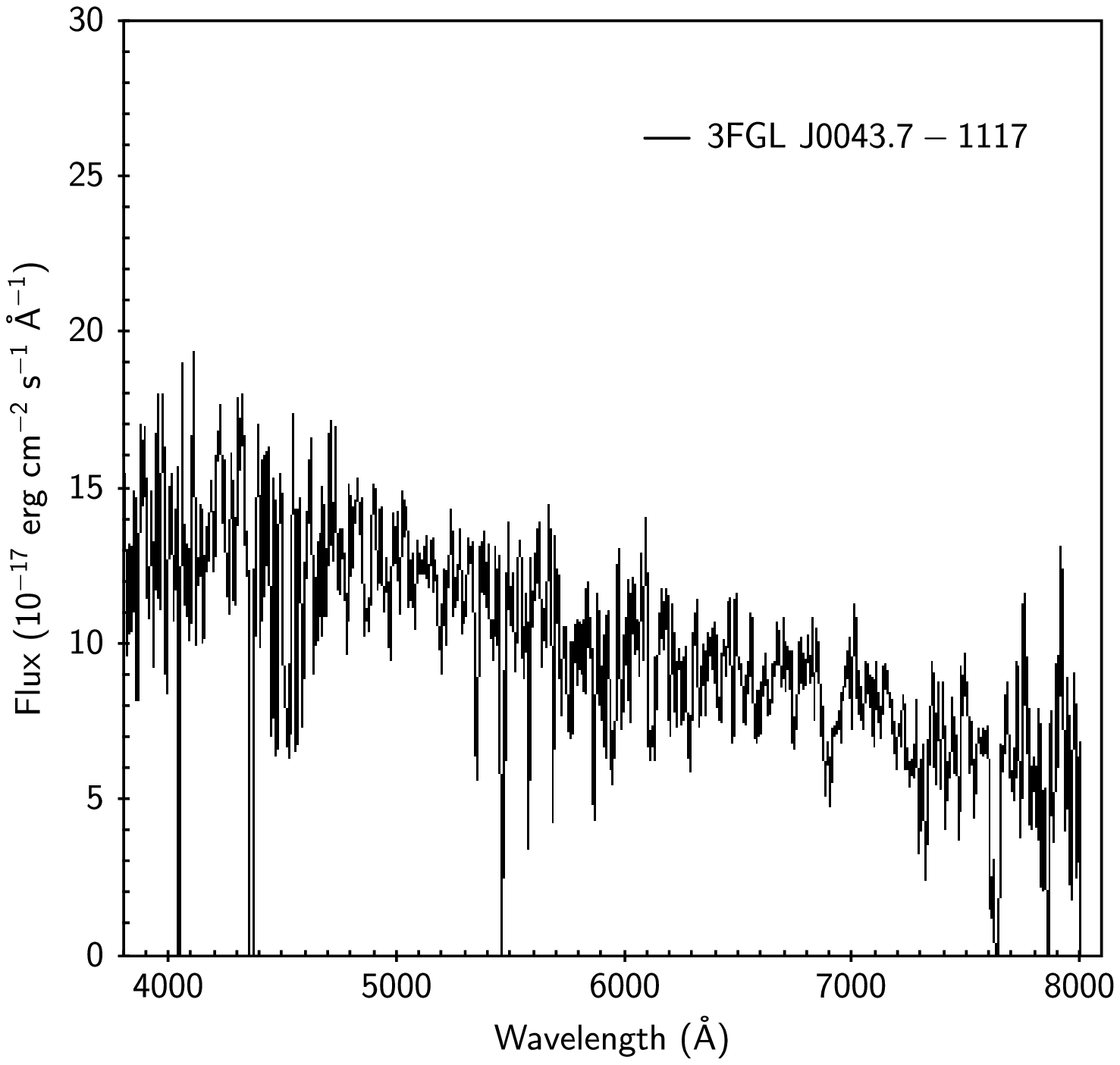}
    \includegraphics[width=0.48\textwidth]{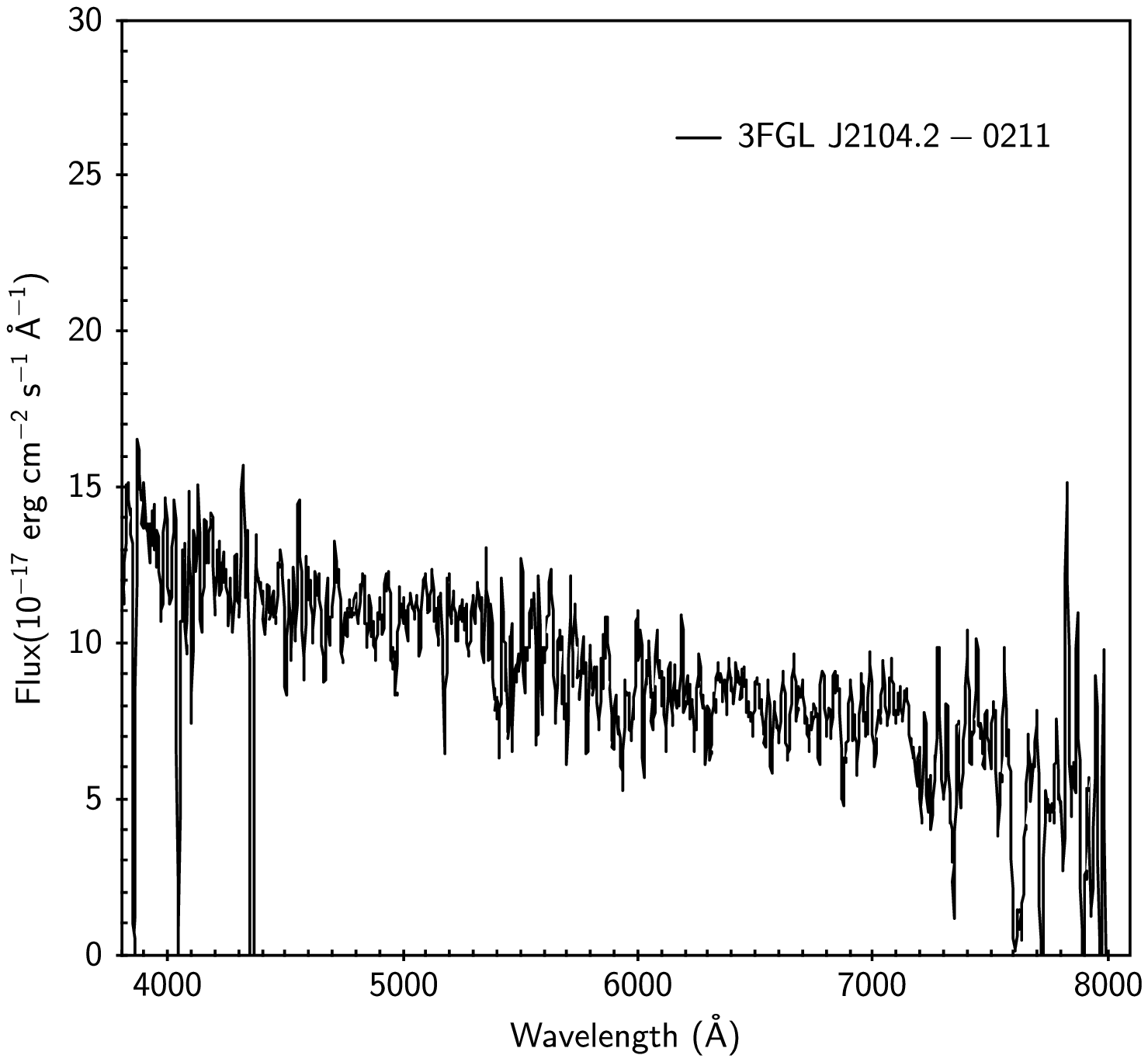}
    \includegraphics[width=0.48\textwidth]{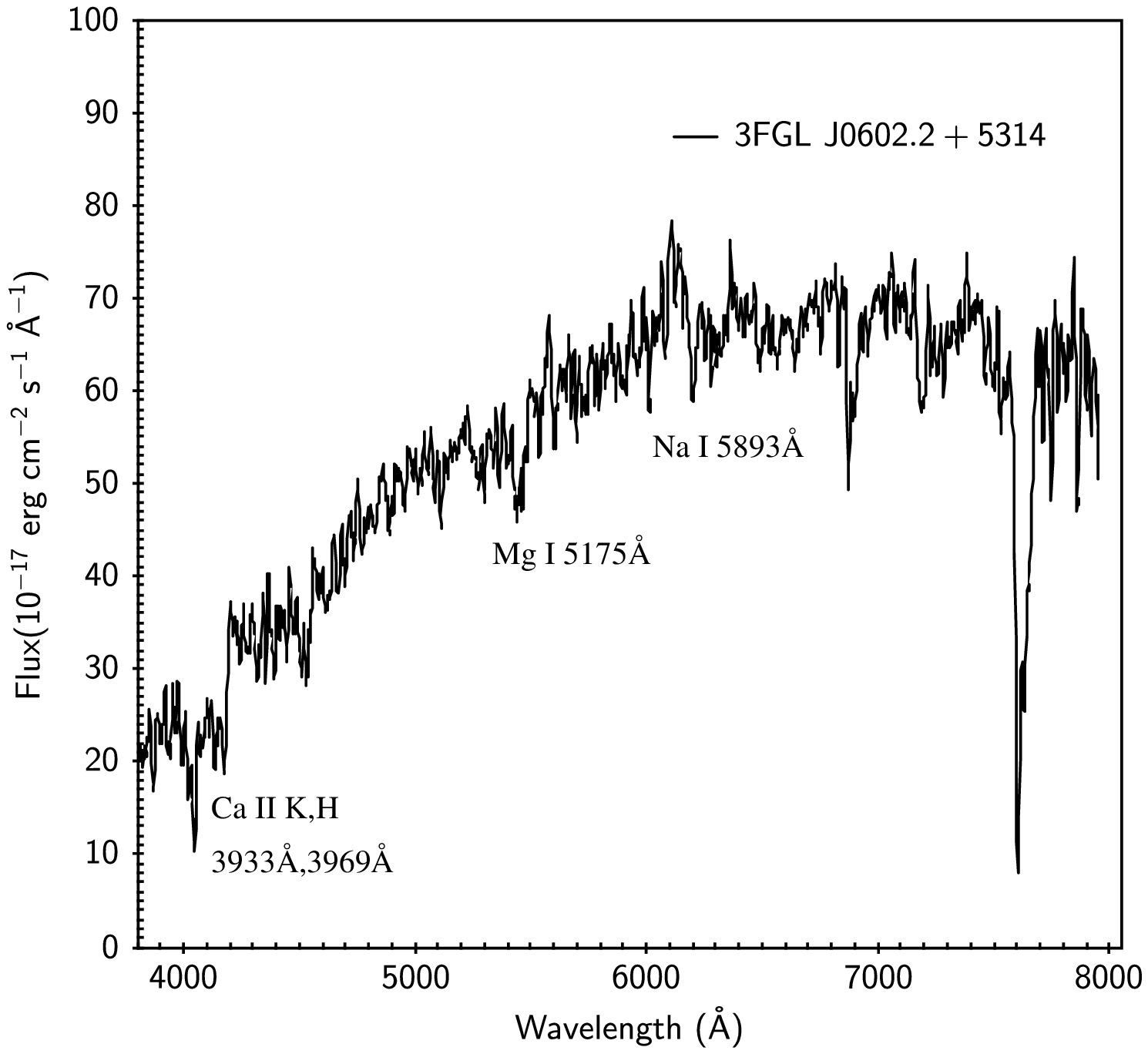}
    \includegraphics[width=0.48\textwidth]{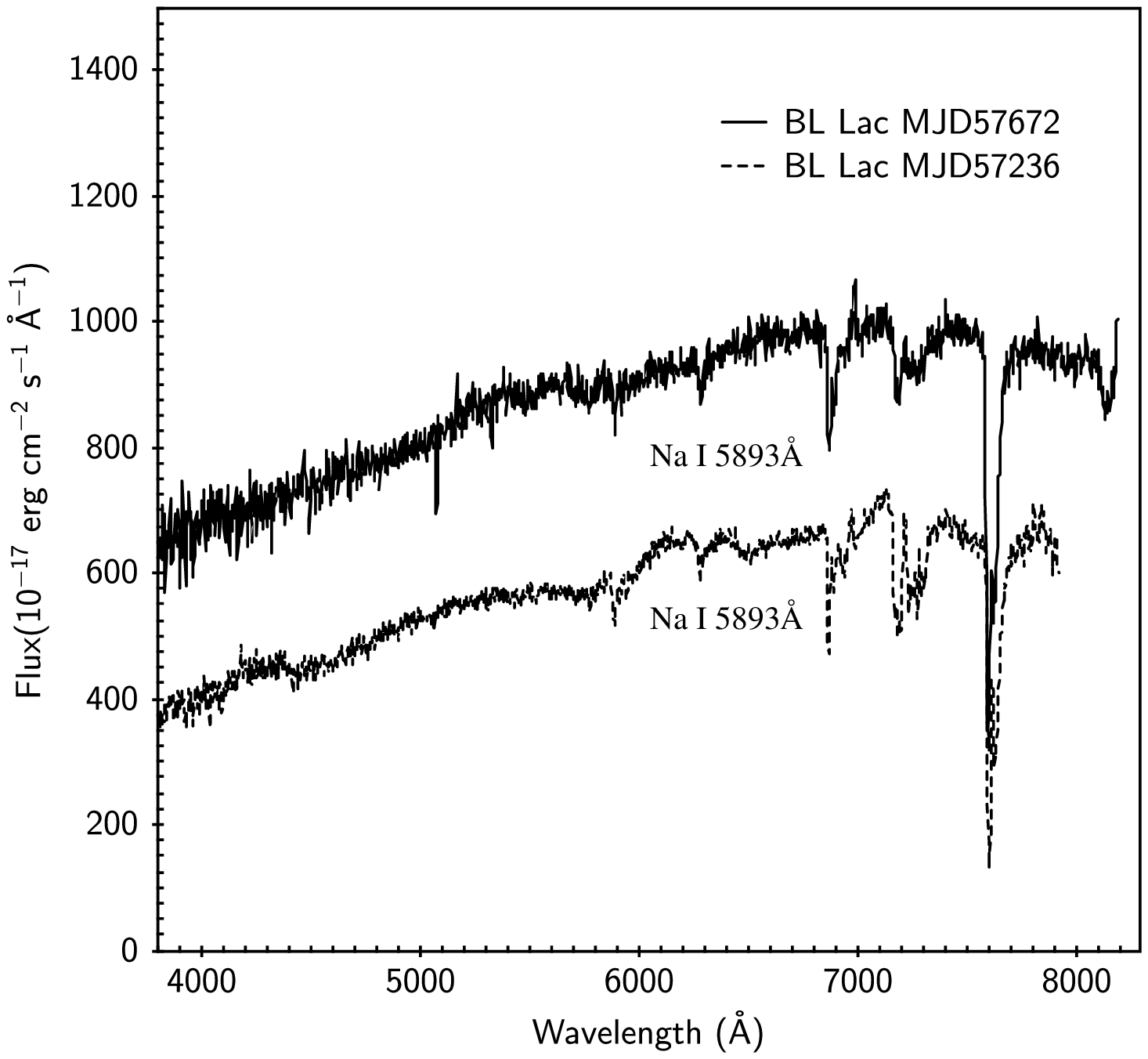}
  \end{center}
  \caption{Spectra of 3FGL~J0043.7-1117 (upper left panel), 3FGL~J2104.2-0211 (upper right panel), 3FGL~J0602.2+5314 (lower left panel) and BL Lac, on 2015 Aug., 2 (dotted line in lower right panel) and on 2016 Oct., 11 (continuous line in lower right panel). The first two objects are Fermi/LAT detected BCUs, having optical spectra with the featureless power-law continuum characteristic of the BL Lac type. The third is a BCU, featuring an optical spectrum of elliptical galaxy at $z = 0.052$, very similar to BL Lac in a low activity state. The class prototype is illustrated, for comparison, in two different activity states, where the stronger power-law contribution of the high activity state can be clearly appreciated.
\label{Spec_Asiago}}
\end{figure*}

With our systematic search of low energy counterparts to the $\gamma$-ray AGNs that are expected to host relativistic jets, we try to overtake the problem both by extending the sample of known objects, as well as by providing additional coverage to the multiple-wavelength programs that monitor AGN jet activity, with the aim to constrain the origin of the various radiation components. In this specific case, the spectra illustrated in Fig.~\ref{Spec_Asiago} are particularly suggestive. Together with sources dominated by the classic power-law contribution of blazars, we note the existence of objects, associated with X-ray emission and flat spectrum radio sources, which produce nearly normal elliptical galaxy spectra in the optical domain. Object of this kind have already been reported by \cite{LaMura15} to have multiple frequency SEDs that closely resemble those of blazars, with the exception of a IR / optical excess, related to the host galaxy star light emission. This kind of sources is very likely hosting a faint jet activity that cannot clearly emerge from the surrounding environment. The idea that a faint jet activity, leading to the emission of a weak power-law continuum, might actually play a role in some objects is supported by the monitoring observations executed on BL Lac itself: comparing the optical spectra obtained in Asiago on August 2$^\mathrm{nd}$, 2015 and on October 11$^\mathrm{th}$, 2016 (this last one as a response to an outburst trigger alert \cite{Atel9599}), we observe a clear higher flux state of the power law spectrum of the outburst period, with respect to the previous state, in which the measured flux was lower and the absorption features of the spectrum were more pronounced.

\section{Conclusions}
\label{secConclusion}
In this work, we considered some fundamental aspects of the theory of radiation emission from relativistic AGN jets and we compared the expected implications with the latest observational data, obtained from the Fermi/LAT $\gamma$-ray monitoring program and from subsequent multiple-wavelength investigations and optical spectroscopic follow-ups. The results of the analysis agree very well with the expectations predicted by a jet radiation model based on synchrotron emission and inverse Compton scattering off ultra-relativistic particles. Many questions, concerning the actual jet composition, its acceleration, the structure of the involved magnetic fields and the nature of the low energy seed photons for the scattering process, however, still have to be properly answered.

Although the general features of $\gamma$-ray emission from jets fit well in the theoretical model that has been nowadays sketched, some fundamental details, like the site of $\gamma$-ray production itself, are still unclear and they point towards different and sometimes contradictory evidence in various objects or within the same object at different epochs. Variability of emission lines, corresponding to $\gamma$-ray flaring activity and ejection of super-luminal components from radio cores has been reported, suggesting that the non-thermal radiation of the jet might contribute to the ionization of the BLR or even produce a BLR-like structure at large distances from the AGN centre \cite{LeonTavares13}. A more extensive investigation of the broad lines, however, indicates that the contribution can at most only partially account for the observed line fluxes, which still are affected by the thermal ionizing radiation of the accretion field \cite{Isler13}. All such problems, in addition, are more likely to affect powerful FSRQs and AGNs where a prominent thermal ionizing radiation component is expected to produce a BLR with a sufficiently strong optical/UV radiation field. In the case when this component is not observed, such as in some BL Lac type objects and in the low luminosity AGN that might be powered by an Advection Dominated Accretion Flow (ADAF), the source of seed photons for the production of $\gamma$-rays and its location along the jet is far less understood. 

Optical spectroscopic investigation, combined with multiple frequency analysis, especially of variable targets of different types, may eventually lead to some insight on the distribution of the different radiation emission sites at various frequencies in the jets, therefore helping to solve such problems. Based on our analysis and on results presented in the extensive literature related to this topic, we can conclude that AGN jets may not share exactly the same nature in all sources, or even at different scales in the same source. The observational hints pointing towards the existence of low power jets, which might not be able to emerge above the host galaxy light, in particular, are suggestive of important effects. Depending on the jet power, indeed, substantial differences may exist between various objects or moving along the jet. The properties of the relativistic blobs of plasma, which are often observed to move at apparently super-luminal speeds through Very Long Baseline Interferometric systems (VLBI), are typically better explained in terms of pair plasmas. Moreover, it is generally expected that a substantial amount of $e^\pm$ pairs can be produced in the vicinity of SMBHs, where jets should be first accelerated. Considerations on the energy loss rate, combined with the estimated kinetic power of jets, however, lead to immediate problems in explaining the large scale extension of jets, which, therefore, require additional contributors to carry their energy. A mixture of hadronic components and of the Poynting flux associated to the extremely well collimated high energy radiation fields, produced within deeper regions of the jet stream, can potentially solve the problem. Some support for this idea stems from the evidence that high energy $\gamma$-rays can be produced in various sites within the jet (e. g. \cite{Finke16}). If this is true, the bulk of the jet radiation might be produced in sites of matter acceleration, either in a shock front or in a region were part of the collimated primary radiation beam is converted into kinetic energy of material flows, which become sources of lower frequency emission.

All of the above mentioned processes, together with the jet interaction with the host galaxy environment, can play a relevant role in determining the jet structure and its extension. A better understanding of what key processes affect the actual properties of relativistic AGN jets will certainly emerge when high resolution studies of fundamental radiation properties, like the degree and the direction of polarization, will be available to perform multiple wavelength radiation variability analysis also in the high energy domain.

\begin{acknowledgement}
GLM gratefully thanks the LOC and SOC of the XXVIII SPIG Symposium for the invitation to contribute an invited topical lecture. This work has been supported by funding from the AstroMundus Master Degree Programme. The authors would like to thank the anonymous referee for suggestions leading to the improvement of this paper.

The Fermi-LAT Collaboration acknowledges support for LAT development, operation and data analysis from NASA and DOE (United States), CEA / Irfu and IN2P3 / CNRS (France), ASI and INFN (Italy), MEXT, KEK, and JAXA (Japan), and the K.A. Wallenberg Foundation, the Swedish Research Council and the National Space Board (Sweden). Science analysis support in the operations phase from INAF (Italy) and CNES (France) is also gratefully acknowledged.

Funding for the Sloan Digital Sky Survey IV has been provided by the Alfred P. Sloan Foundation, the U.S. Department of Energy Office of Science, and the Participating Institutions. SDSS-IV acknowledges support and resources from the Center for High-Performance Computing at the University of Utah. The SDSS web site is www.sdss.org.

SDSS-IV is managed by the Astrophysical Research Consortium for the Participating Institutions of the SDSS Collaboration including the Brazilian Participation Group, the Carnegie Institution for Science, Carnegie Mellon University, the Chilean Participation Group, the French Participation Group, Harvard-Smithsonian Center for Astrophysics, Instituto de Astrof\'isica de Canarias, The Johns Hopkins University, Kavli Institute for the Physics and Mathematics of the Universe (IPMU) / University of Tokyo, Lawrence Berkeley National Laboratory, Leibniz Institut f\"ur Astrophysik Potsdam (AIP),  Max-Planck-Institut f\"ur Astronomie (MPIA Heidelberg), Max-Planck-Institut f\"ur Astrophysik (MPA Garching), Max-Planck-Institut f\"ur Extraterrestrische Physik (MPE), National Astronomical Observatory of China, New Mexico State University, New York University, University of Notre Dame, Observat\'ario Nacional / MCTI, The Ohio State University, Pennsylvania State University, Shanghai Astronomical Observatory, United Kingdom Participation Group, Universidad Nacional Aut\'onoma de M\'exico, University of Arizona, University of Colorado Boulder, University of Oxford, University of Portsmouth, University of Utah, University of Virginia, University of Washington, University of Wisconsin, Vanderbilt University, and Yale University.

This work is based on observations collected at Copernico telescope (Asiago, Italy) of the INAF - Osservatorio Astronomico di Padova and on observations collected with the 1.22m \textit{Galileo} telescope of the Asiago Astrophysical Observatory, operated by the Department of Physics and Astronomy "G. Galilei" of the University of Padova.
\end{acknowledgement}

{\noindent \bf Author contributions}\\
G. La Mura: speaker, author of text, designer of the multiple frequency source identification procedure.\\
G. Busetto: physics advisor for the theory of radiation from relativistic charged particles.\\
S. Ciroi: data reduction and analysis specialist.\\
P. Rafanelli: scientific advisor of the multiple frequency AGN investigation program.\\
M. Berton, E. Congiu, V. Cracco, M. Frezzato: optical counterpart monitoring campaign observers.\\

\end{document}